# Network-Based Subbasin-Scale Mapping of Streamflow Alteration in Ontario, Canada

Hongren Shen[1,*], Bryan A. Tolson[1], James R. Craig[1], Robert A. Metcalfe[2,3], Jonathan Romero-Cuellar[1], and James J. Luce[3]

[1]Department of Civil and Environmental Engineering, University of Waterloo, Ontario, Canada
[2]Environmental and Life Sciences Graduate Program, Trent University, Ontario, Canada
[3]Aquatic Research and Monitoring Section, Ministry of Natural Resources, Trent University, Ontario, Canada

[*]*Correspondence to*: hongren.shen@uwaterloo.ca

**Abstract.** Dams, reservoirs, and waterpower facilities are central to regional water management, but they can induce streamflow alterations that propagate downstream through connected river networks and complicate hydrologic modeling and aquatic ecosystem assessment. Existing streamflow indicators often quantify alteration only at discrete locations (e.g., hydrometric gauges), thus limiting spatially continuous, subbasin-scale mapping of where upstream infrastructure may influence downstream flows. We propose the Streamflow Alteration Index (SAI), a network-based screening metric that propagates point-based alteration signal sources downstream through a routing network to map potential alteration influence. The framework includes paired indices: SAI_I (0–100%) and SAI_II (0–100%), representing two levels of influence derived from confirmed (Level I) and potential (Level II) alteration signal sources. We implemented SAI to develop the SAI v1.0 database for Ontario, Canada, using the Ontario Lake and River Routing Product Version 2 (OLRRP v2) network and an inventory of 643 alteration signal sources compiled from provincial datasets. The resulting product provides seamless, high-resolution, network-based, subbasin-scale mapping of streamflow alteration across Ontario and enables SAI estimates at both gauged and ungauged nodes within the routing network. We validated SAI by classifying subbasins containing hydrometric gauges as natural, conditional, or altered and compared these classes against two independent references: Reference Hydrometric Basin Network (RHBN) gauges (with minimal human impact) and Water Survey of Canada (WSC) gauge "Natural" and "Regulated" labels. SAI shows strong agreement for gauges labelled natural: 86% of RHBN natural gauges and 86% of WSC-natural gauges are classified as natural by SAI. For WSC-regulated gauges, SAI classifies 74% as conditional or natural. Overall, SAI v1.0 provides a practical, objective screening product for large-domain hydrologic and aquatic applications without requiring detailed dam operating information, and it offers a scalable pathway toward national- to global-scale, network-based screening products for streamflow alteration.

## 1. Introduction

Dams and reservoirs are among the most significant infrastructural interventions in freshwater systems, constructed to serve a range of human needs including water supply, flood and drought control, irrigation, waterpower generation, and recreation (Maavara et al., 2020; Palmer and Ruhi, 2019; WCD, 2000). As documented in the Global Dam Watch database (Lehner et al., 2024), there are 41,145 barrier locations and 35,295 associated reservoirs globally, with a cumulative storage volume of 7,420 km$^3$ and a surface water area of 304,600 km$^2$. More broadly, river networks worldwide are affected by many more instream structures, and the majority of large rivers worldwide are fragmented and regulated (Grill et al., 2019). Engineered watersheds have played a crucial role in enabling agricultural expansion, supporting urban water demands, mitigating natural disasters, and driving economic development (Lehner et al., 2011; WCD, 2000).

However, the widespread construction and operation of dams have profoundly transformed natural flow regimes, altered sediment and nutrient fluxes, and generated substantial ecological and social trade-offs (Richter et al., 2010; Winemiller et al., 2016). Dams and reservoirs can substantially modify the magnitude, timing, frequency, and duration of natural streamflows by regulating the storage and release of water. Such streamflow alteration is widely regarded as a first-order impact on fluvial systems, because flow regimes are a fundamental driver of riverine habitat and ecosystem processes (Burke et al., 2009; Jorde et al., 2007; Petts, 1984; Poff et al., 1997, 2010). In this study, streamflow alteration denotes human-caused changes to natural streamflow patterns arising from instream structures and lake outflow controls, primarily through the construction and operation of dams, reservoirs, and waterpower facilities. Such alteration commonly dampens high flows, augments baseflows, and smooths natural variability, with widespread implications for river connectivity, ecosystem dynamics, and water resources management. Hereafter, the term alteration is used throughout, except where regulation is retained for consistency with established terminology (e.g., degree of regulation, HYDAT natural and regulated labels) and cited sources.

Streamflow alterations occur across temporal scales ranging from minutes to seasons, with particularly strong sub-daily variability downstream of hydropeaking waterpower facilities (WPFs). Hydropeaking refers to sub-daily operational changes in discharge used to match electricity generation to demand, which commonly produces higher releases during daytime, when electricity demand is greatest, and lower releases at night. This creates highly variable downstream hydraulic and habitat conditions at short-time scales (Bejarano et al., 2018; Kern et al., 2012). Because these fluctuations happen on sub-daily time scales, daily discharge records can mask key features of alteration such as ramping rates, the frequency of peaks, and the duration of high- and low-flow periods (Smokorowski, 2022). At longer (seasonal-to-annual) scales, reservoir operations commonly suppress flood peaks during wet periods and augment low flows during dry periods, which partially decouples discharge from natural hydroclimatic variability (Déry et al., 2016). The magnitude and spatial footprint of these alterations depend on interacting factors, including project purpose (e.g., flood control, waterpower, recreation), reservoir storage and dam operating rules, river-network configuration (e.g., reservoir size relative to inflow, cascades of dams and diversions, and channel hydraulics), and basin physiography and hydroclimate (Burke et al., 2009; Jorde et al., 2007).

A recent systematic review of 155 studies by Bipa et al. (2024) further showed that hydropeaking affects more than discharge alone: it can also modify water temperature, sediment processes, and dissolved gas conditions, with well-documented consequences for fish (e.g., movement, spawning, migration), benthic macroinvertebrates, and riparian vegetation. Global reliance on dams is expected to remain high or even increase in response to growing water stress, energy demand, and climate change, making robust assessment of dam impacts increasingly important (Steyaert et al., 2025; Tharme, 2003).

In the hydrologic modeling community, where model implementations can often assume natural conditions, streamflow alteration can negatively influence model calibration and validation. Symptoms include biased parameter estimation, reduced model transferability, and unreliable predictions under changing conditions. These impacts are especially pronounced in large-scale modeling, where reservoir operations and other human-water processes are difficult to represent consistently due to data limitations, simplified operating rules, and challenges in evaluation across large domains (Dang et al., 2020; Shrestha et al., 2024; Wada et al., 2017; Yassin et al., 2019; Zajac et al., 2017).

Assessing flow alteration is increasingly important for hydrologic modeling and water management purposes, and it is critical for both process-based and data-driven models. Altered flow regimes remain difficult to represent in process-based hydrologic models, especially at large scales where reservoir operations are often simplified and poorly constrained by observations (Wada et al., 2017). Recent deep-learning rainfall-runoff studies showed that conditioning models on static catchment attributes substantially improves generalization across basins by helping networks learn basin-to-basin differences in storage and response behavior (Kratzert et al., 2019a, b). However, in human-altered rivers, discharge reflects not only hydroclimate and physiography but also operational decisions; incorporating regulation descriptors as attributes has been shown to improve prediction skill in regulated catchments (Tursun et al., 2024). Moreover, alterations can undermine parameter transfer in regionalization by breaking hydrologic similarity assumptions, thus complicating streamflow prediction in poorly-gauged or ungauged basins (Blöschl, 2013; Dang et al., 2020; Du et al., 2022; Hrachowitz et al., 2013; Montanari et al., 2013). This need for scalable assessment is also consistent with regional environmental-flow frameworks that recognize the impracticality of evaluating hydrologic alteration on a river-by-river basis across large domains (Poff et al., 2010).

The impacts of dam and reservoir-caused alteration on hydrology have been widely documented in recent decades (Déry et al., 2016; Grill et al., 2014; Habets et al., 2018; Lehner et al., 2011; Richter et al., 2010; Yang et al., 2022). In parallel, numerous indices have been developed to quantify regulation intensity and related impacts across river systems, ranging from storage-based metrics (e.g., degree of regulation) to fragmentation and connectivity metrics used in large-scale assessments (e.g., Nilsson et al., 2005; Grill et al., 2014), including reach-based, decadal dam impact metrics that propagate cumulative upstream storage effects through river networks (e.g., Wieczorek et al., 2021).

One of the most widely used storage-based measures is the degree of regulation (DOR), a residence time proxy defined as the ratio of the total reservoir storage capacity to the mean annual flow. Higher DOR values indicate greater potential for regulation influence. The DOR has been mapped globally and is included as an attribute in several global dam/reservoir datasets (Lehner et al., 2011, 2024; Nilsson et al., 2005; Yin et al., 2024). However, like all metrics, DOR has several limitations: (1) using total reservoir storage capacity (both active and dead storage) can overestimate regulation impacts (Kumar et al., 2022; Lehner et al., 2011); (2) global DOR estimates often rely on empirically inferred storage capacities and modelled mean annual river flow, which introduce substantial uncertainty (Lehner et al., 2011); (3) reservoir-by-reservoir DOR does not adequately represent cumulative effects of cascading reservoirs within a river network (Habets et al., 2018; Kumar et al., 2022); and (4) DOR is defined at a dam/reservoir and does not directly propagate downstream. Overall, these limitations make DOR less suitable for subbasin-scale hydrologic applications that require cumulative network effects, downstream propagation beyond the regulation sources, and strong grounding in local inventories to produce reliable spatial estimates of streamflow alteration.

Another commonly used concept for characterizing dam and reservoir impacts is the zone of influence (ZOI), which delineates the upstream and downstream river-network areas affected by a dam, reservoir, or WPF (SEI International, 2009; Stainton, 2012). The hydrologic ZOI provides a practical framing of the spatial footprint of flow regulation beyond the structure itself. The ZOI concept is also relevant in broader river management contexts, since flow alteration footprints can inform aquatic ecosystem assessment and impact evaluation, although the ecological consequences of regulation are not necessarily bounded by simple hydrologic distance- or area-based rules (Metcalfe et al., 2013).

Stainton (2012) noted that the upstream ZOI is often straightforward to delineate (e.g., reservoir inundation), whereas the downstream ZOI is more difficult to define because it lacks a clear boundary and depends on operating regime, river hydraulics, and basin context. In this framing, the downstream ZOI extends from the tailrace to the point where operationally induced alterations become sufficiently attenuated or diluted that they are no longer discernible from background conditions. Stainton (2012) reviewed simple drainage area-based prediction rules, including a catchment-ratio threshold adapted from Petts (1978, 1979, 1980) and the British Columbia "area of impact" multiplier (Lewis et al., 2004), but showed that these heuristics can substantially under- or over-estimate influence when tested in Ontario basins. These studies highlight both the usefulness and the limitations of simple drainage area-based heuristics for delineating downstream alteration footprints and point to the need for approaches that more explicitly represent cumulative network effects and are better suited to large-sample hydrologic applications.

The decadal dam impact (DI) metric of Wieczorek et al. (2021) characterizes the degree of dam regulation for individual river reaches over time based on cumulative upstream reservoir storage, the relative drainage area represented by upstream dams, and the volume of precipitation within the reach drainage area. Unlike reservoir-level indices, DI is defined at the river-reach scale and explicitly represents cumulative upstream effects through the river network. This reach-based, temporally structured formulation is a major strength, as it moves beyond single-reservoir metrics and enables large-scale assessment of potential cumulative dam influence. However, DI remains fundamentally dependent on reservoir storage information and thus inherits limitations common to storage-based approaches, including uncertainty in reported or inferred storage capacity and the fact that storage alone cannot fully represent reservoir operations or actual streamflow alteration. Although DI is available for millions of river reaches across the conterminous United States (US), its transferability to Canada and other regions is constrained by incomplete reservoir storage information and by the current US-only coverage of the published dataset.

To better quantify and map the impacts of streamflow alteration across basin drainage networks in Canada, we introduce a new, relatively simple streamflow alteration metric: the Streamflow Alteration Index (SAI). SAI represents the potential influence of dams and reservoirs at any location in a basin and supports spatially explicit mapping of streamflow alteration. Specifically, SAI is defined as the percentage of upstream drainage area that is routed through an alteration source and, by extension, provides a proxy for the fraction of streamflow at that point that passes through an upstream alteration source. By propagating alteration signals downstream through the river network, SAI addresses a key limitation of many traditional metrics that focus on individual dams or reservoirs and do not explicitly represent influence moving downstream from the alteration sources. Moreover, SAI is designed to be practical in data-limited settings: unlike storage- or discharge-based indices (e.g., DOR and DI), it does not require detailed time-varying information (e.g., long-term streamflow or precipitation records) or hard-to-estimate infrastructure attributes (e.g., active reservoir storage). Instead, SAI relies on identifying a key static attribute—whether a dam or reservoir is expected to substantially modify downstream flows—and then propagating that influence through the river network. This design allows consistent application across large domains where operational and storage data are incomplete or unavailable. We developed the SAI v1.0 database for Ontario, Canada, leveraging the high-resolution Ontario Lake and River Routing Product Version 2 (OLRRP v2), which discretizes Ontario into nearly 250,000 subbasins with an average area of 4.3 km$^2$ (Shen et al., 2024). The database provides subbasin-scale SAI maps and associated GIS layers that support the classification of hydrological subbasins and hydrometric gauges and include harmonized attributes describing alteration signal sources and gauge metadata.

It should be noted that the present SAI framework focuses on flow alteration arising from instream structures and lake outflow control structures, including dams, reservoirs, waterpower facilities, and lockstations, that modify water levels, storage, and releases. Accordingly, it does not explicitly account for other anthropogenic influences on streamflow, such as wastewater treatment plant effluent, industrial discharges, irrigation withdrawals and return flows, managed aquifer storage and recovery, landscape change (e.g., urbanization), or inter-basin transfers and canal or diversion systems. In many basins, these influences are secondary relative to large dams, but they can still represent a non-trivial fraction of discharge during low-flow periods and may meaningfully affect flow magnitude and seasonality. These mechanisms are not incorporated here because consistent province-wide datasets are not available for all source types. Nevertheless, the SAI framework could be extended to include them in future applications where suitable databases and network attribution become available.

The remainder of this paper is structured as follows. Section 2 details the data sources and methodology for the development of the SAI v1.0 database for Ontario, as well as the approach to validating the database. Section 3 presents the SAI v1.0 database and the validation results. Section 4 discusses the intended use, limitations, and broader applicability of the dataset. Section 5 summarizes the main conclusions.

## 2. Materials and Methods

The impacts of streamflow alteration caused by dams and reservoirs manifest spatially rather than being confined to the point scale. Impacts generally diminish downstream from each alteration signal (or site) due to the mixing of the altered flow signals with the signal from natural incremental runoff. However, if a new alteration site is introduced along the way, the impacts may be amplified again. Therefore, identifying all likely sources of instream alteration is critical in mapping their impacts on streamflow in space.

In this section, the data and methodology used to develop the Streamflow Alteration Index (SAI) are presented. Section 2.1 describes main data sources for identifying alteration signals and validating SAI results. Section 2.2 defines alteration signals and summarizes how they are identified from different data sources. Section 2.3 introduces how alteration signal points are snapped to the river network, which is the key data processing procedure in SAI development. Section 2.4 demonstrates the SAI calculation procedures. Section 2.5 describes how SAI results are validated against locations with reported regulation conditions.

### 2.1 Data Sources

Six main datasets are used to develop and validate the SAI database for Ontario:

- **Ontario Lake and River Routing Product Version 2 (OLRRP v2):** The OLRRP v2 is a high-resolution, vector-based lake and river routing GIS product over the Province of Ontario (Shen et al., 2024) and is an update to OLRRP v1 introduced by Han et al. (2023). It supplies network topology and various subbasin, channel, and lake attributes required for hydrological modeling. The OLRRP v2 connects 245,576 subbasins and 82,928 lakes (surface areas ranging from 0.1 $km^2$ to 4,506 $km^2$) and assigns 4,033 points of interest (POI) to these subbasins and lakes. Lake polygons and attributes were employed from the HydroLAKES database (Messager et al., 2016). Each POI was treated as a pour point during the delineation process, ensuring accurate placement at subbasin outlets. These POI were derived from multiple point-based datasets, including hydrometric monitoring gauges from the Water Survey of Canada (WSC). The OLRRP v2 demonstrated high spatial accuracy for POI locations, with 74% of WSC gauges showing less than 2% difference between their delineated and documented drainage areas. Only 4% of WSC gauges exhibit absolute drainage area differences greater than 10% (Shen et al., 2024). Overall, this product provides a high-resolution lake and river routing network for SAI development.
- **Ontario dams:** Three provincial dam databases provided by Ontario Ministry of Natural Resources (MNR) are used: (1) The Ontario Dam Inventory (ODI), which is a point-based inventory of medium and large dams throughout Ontario with dam names and dam owner information provided (Ontario Ministry of Natural Resources, 2014). We used the version accessed in January 2024. The ODI contains 1,839 dam points owned and managed by a variety of entities, including government agencies at federal, provincial, and municipal levels, conservation authorities, indigenous communities, private owners, and Ontario Power Generation; and (2) a dataset of 389 Ontario MNR-operated dams that contains two additional attributes, including dam operation status (operational or decommissioned) and dam purposes (e.g., flood control, low flow augmentation, navigation, and recreation). Note that these two datasets have 347 dam points overlapping, and the total net count is 1,881.
- **Waterpower facilities (WPFs).** We used the Waterpower Generation Stations dataset maintained by the Ontario MNR (retrieved the version published in December 2024) to characterize potential streamflow alteration associated with waterpower infrastructure. For all retrieved 211 stations, the dataset provides an operation type (OPERATION_TYPE), which distinguishes run-of-river facilities from plants designed for load-following (e.g., peaking, intermediate) or pumped-storage operation. This field offers a descriptor of how a facility is typically operated and the management objectives that may drive water level and release decisions. In addition, we used the river system plan type provided in the Ontario provincial Water Management Plan (WMP) (simple vs. complex; Ministry of Natural Resources (2016)) as a third, system-level indicator of operational context. This attribute is originally associated with rivers and WPFs, which reflects whether river operations are managed under limited control (simple) or under more coordinated, multi-structure planning (complex).
- **Lockstations.** Lockstations were incorporated to represent flow regulation associated with navigational water-level management in Ontario. We digitized/compiled 72 lockstations located on the Trent-Severn Waterway, the Rideau Canal, and the Welland Canal, which collectively include the lock systems in the Trent River-Rice Lake-Kawartha Lakes-Lake Simcoe corridor, along the Rideau Canal, and along the Welland Canal between Lake Ontario and Lake Erie. Lockstation locations and supporting site information were obtained from the official Parks Canada lockstation directories and associated lockstation pages for the Trent-Severn Waterway and Rideau Canal, and from official Great Lakes St. Lawrence Seaway sources for the Welland Canal locks. These lockstations were treated as point-based infrastructure that can alter downstream streamflow through operational control structures (e.g., lock gates and related water-control works) and were integrated into the routing network as part of the SAI v1.0 alteration signal inventory.

- **Hydrometric Gauges:** We compiled 1,015 hydrometric gauges from the Water Survey of Canada (WSC) that are embedded in the OLRRP v2 POI layer. Many of these gauges carry a WSC regulated/natural flag (Water Survey of Canada, 2024). However, the criteria and rationale used to assign these classifications are not fully documented, which can introduce uncertainty in analyses that distinguish regulated from natural conditions using WSC flags alone. Despite this limitation, the WSC regulation flag remains widely used in Canadian hydrological studies since it is currently the only publicly available, nationwide indicator of regulation status (Arsenault et al., 2020; de Rham et al., 2020). Here, we compare the WSC binary flag with SAI-derived conditions at gauge locations to illustrate the added value of a spatially explicit, network-propagated streamflow alteration metric.
- **Reference Hydrometric Basin Network (RHBN):** The RHBN is a subset of Canada's national hydrometric monitoring stations, maintained by the WSC. It features high-quality, long-term data from catchments with minimal human impacts, which are essential for understanding natural hydrological variability and serve as a benchmark for climate-related studies (Brimley et al., 1999; Ouarda et al., 1999). The RHBN, updated in 2020, includes 1027 stations across Canada, of which 154 stations are within Ontario (Pellerin, 2019). These 154 stations are used as the primary reference to validate the proposed SAI database in this study.

### 2.2 Identification of Alteration Signal Sources

Alteration signal sources refer to point-based locations where streamflow is first modified by instream infrastructure, represented here by the outlet location of a dam, reservoir, waterpower facility, or lockstation. It should be noted that not all structures captured in available inventories meaningfully alter downstream streamflow. For example, small barriers (e.g., weirs) constructed primarily for local water-level augmentation have limited influence on discharge compared with actively managed reservoirs (e.g., facilities operated for flood control or waterpower generation). Consequently, a key step in developing the SAI database is to filter out sources that are likely to have minor impacts and retain those expected to produce substantial downstream alteration.

At the provincial scale, however, detailed operating rules and regulatory planning information are not consistently available for the thousands of structures managed by diverse entities (e.g., federal and provincial agencies, conservation authorities, the Ontario Ministry of Natural Resources, and private owners). Our assessment of minor versus major alteration therefore relies on information that is broadly available and consistently documented across datasets. We classified each candidate alteration signal source into one of three levels: confirmed (Level I), potential (Level II), and negligible (Level III). Level I sources are identified with high confidence as substantially modifying downstream streamflow. Level II sources are likely to affect streamflow but with greater uncertainty in the presence and magnitude of alteration, and a smaller expected influence than Level I. Level III sources are assumed unlikely to meaningfully alter streamflow and are treated as negligible or insignificant alteration sources for SAI mapping. These categories are summarized in Table 1, and Level I and Level II signals are described further below. In total, we identified 643 Level I and Level II signal sources from 2,475 candidates, where multiple signal sources within the same subbasin are counted once.

- **Confirmed alteration signal sources (Level I)**. Level I signals represent locations where streamflow alteration is expected to be substantial and is supported by strong evidence. For waterpower facilities (WPFs), we classify a facility as a Level I alteration source based on three operational indicators: (1) association with a complex Water Management Plan (WMP), which is typically developed for river systems with multiple WPFs and/or other control structures where coordinated operations exert substantial control over water levels and releases (Ministry of Natural Resources, 2016); (2) an operation type reported as pumped-storage, peaking, or intermediate; or (3) a reservoir purpose reported as waterpower. Moreover, all lockstations are treated as Level I signals, as they are operated as water control structures for navigation and water level management.

- **Potential alteration signal source (Level II).** Since complete and consistently documented information on the operational significance of individual control structures is rarely available at large scales (unavailable in Canada), we define Level II sources as features that are likely to influence streamflow but for which the magnitude and persistence of alteration are more uncertain, but likely less, than for Level I sources. Level II sources are identified using three complementary criteria: (1) All remaining WPFs in the provincial dataset that are not classified as Level I (i.e., facilities lacking strong indicators of substantial alteration but still plausibly capable of affecting flows); (2) Structures in the ODI whose metadata suggest control functions, including dams with names containing the keyword "control dam" and dams owned by conservation authorities or Ontario Power Generation (OPG). We include OPG-owned dams as potential alteration sources because OPG operates a large portfolio of waterpower infrastructure in Ontario, and ownership by a major waterpower operator is a practical proxy for the presence of active water-level and release management even when facility-specific operational details are not available; and (3) MNR dams that are operational (excluding decommissioned or non-operational structures) and whose stated purpose includes flood control and/or low flow augmentation. Together, these criteria provide a pragmatic, evidence-based way to capture plausible alteration signals where detailed operational data are unavailable.
- **Negligible alteration sources (Level III).** Level III sources represent inventoried control structures that are unlikely to meaningfully alter streamflow and are therefore treated as negligible alteration in this study. After identifying Level I and Level II alteration signal sources, all remaining structures in the ODI and the MNR dam dataset are assigned to Level III. These features may include small or pass-through structures, localized controls, or infrastructure with limited storage or operational influence, for which available metadata do not indicate active alteration (e.g., flood control, low flow augmentation, or waterpower-related management). Level III sources are retained to document screening decisions and avoid double counting, but they are not used to propagate alteration signals in the SAI calculations.

### 2.3 Geospatial Alignment for Alteration Signal Points

The Streamflow Alteration Index (SAI) is calculated at the OLRRP subbasin scale, with subbasins roughly 4.3 km$^2$ in area on average. Accurate assignment of alteration signal sources, including dams, reservoirs, waterpower facilities and lock stations, to the correct OLRRP subbasins is therefore essential. We refer to this geospatial alignment process as snapping.

Alteration signal sources and hydrometric gauges were initially represented as point features. Since source datasets contain positional uncertainty, raw point locations do not always coincide with the correct rivers, lakes, or subbasin outlets in OLRRP. We aligned these points to the OLRRP network using a two-stage workflow: (1) automated snapping to the most plausible topologically consistent location, followed by (2) detailed manual review. Snapping was constrained by OLRRP lake and subbasin topology, feature names (string matching), and agency-documented drainage areas where available. Manual verification compared the snapped locations against satellite imagery and online basemaps (Google Maps and Esri Basemaps; Google, 2024; Esri, 2024). The final count of points after snapping is summarized in Table 1.

The OLRRP network is pre-delineated and fixed; thus, some points, particularly small run-of-the-river structures, may remain positioned within a subbasin after snapping rather than exactly at the subbasin outlet. This is acceptable for SAI because OLRRP subbasin resolution is quite high and so within-subbasin placement introduces limited spatial errors.

**Table 1.** Definitions and screening criteria used to classify alteration signal sources into confirmed (Level I), potential (Level

II), and negligible (Level III) categories for Streamflow Alteration Index (SAI) development, based on waterpower facility and dam datasets in Ontario. "Number of sources" reports the count of points successfully snapped to the OLRRP river network and retained for analysis; records with large spatial discrepancies after snapping were excluded (see note below table).

| Datasets | Descriptions | Number of Sources | Level I Criterion | Level II Criterion | Level III Criterion |
|---|---|---|---|---|---|
| Waterpower Facilities (WPFs) | Waterpower generation stations obtained from the Ontario MNR. Generation stations contain attributes such as the operation type (e.g., run-of-river, peak load, etc.), associated reservoir purposes (e.g., flood control, navigation, etc.), and water management plan type (complex vs. simple) (Ministry of Natural Resources, 2016) | 211 | Has a complex water management plan; or the operation type is pumped-storage plant, peak load plant, or intermediate plant | All remaining generation stations not included in Level I | \ |
| Ontario Dam Inventory (ODI) | The ODI provides medium to large dam along with attributes including their ownership and a name string across Ontario. | 1803* | \ | Dams containing keyword "control dam" in their dam names; or dams owned by conservation authorities or Ontario Power Generation (OPG) | All remaining ODI not included in Level II |
| MNR Dams | The Ontario MNR operated dams contain additional information not within the ODI such as dam operation status and dam purpose. | 389 | \ | Dams for flood control, and low flow augmentation; and dams are not decommissioned or non-operational | All dams not included in Level II |
| Lockstations | Lockstations from the Trent-Severn Waterway, the Rideau Canal, and the Welland Canal are included as point-based signals representing navigational flow control structures. | 72 | All lockstations | \ | \ |

*The original Ontario Dam Inventory (ODI) includes 1,839 dam points. Of these, 36 were excluded from the alteration signal sources due to substantial spatial mismatches between the reported dam locations and the Ontario Lake and River Routing Product (OLRRP), which could not be resolved using secondary data sources or agency contact.

### 2.4 Streamflow Alteration Index (SAI)

In many studies such as hydrologic modeling, whether a location (e.g., a hydrometric gauge) is affected by upstream alteration is often represented as a binary agency flag (natural vs. regulated), and detailed reconstruction of dam operating rules is uncommon in large scale applications. Although explicit reservoir operating rules can improve model performance in regulated basins (Craig et al., 2020; Hughes et al., 2021; Salwey et al., 2024; Yassin et al., 2019), these data are rarely available consistently at large scales. Such practical constraints motivate spatially explicit, network-aware indicators of alteration that can be derived from limited but widely available information.

Our Streamflow Alteration Index (SAI) quantifies the likely degree of streamflow alteration upstream of any location in a river network (e.g., a subbasin outlet). SAI is an area-based metric that propagates alteration signals through the drainage network, summarizing the percentage of upstream drainage area routed through locations that contain alteration signal sources. Assuming streamflow is proportional to upstream drainage area, a SAI value of 30% at some point on the river network indicates that 30% of the streamflow at that point is routed through at least one upstream alteration source like a dam. We compute SAI at the subbasin scale (at the subbasin outlet points only) to support spatially continuous mapping of dam and reservoir-related alteration across large domains, including ungauged basins.

To reflect differences in confidence and expected influence among alteration sources, we compute two complementary indices: SAI_I, representing alteration associated with confirmed (Level I) sources, and SAI_II, representing alteration associated with potential (Level II) sources. The two indices are calculated independently: SAI_I is derived from Level I sources only and ranges from 0 to 100%, whereas SAI_II is derived from Level II sources only and ranges from 0 to 100%. Figure 1 illustrates the SAI calculations using a hypothetical subbasin-river network. The calculation procedure consists of three steps.

**Step 1: Identify alteration signal sources.** Identify all Level I and Level II alteration signal sources for the basin (subbasins B, E, G and M in Figure 1a). Signal source classification is described in Section 2.2.

**Step 2: Initialize SAI_I and SAI_II.** For each subbasin, initialize the two indices based on the presence of alteration sources within the subbasin. If a subbasin contains one or more Level I sources, set SAI_I = 100%; otherwise set SAI_I = 0. If a subbasin contains one or more Level II sources, set SAI_II = 100%; otherwise set SAI_II = 0. Since the indices are independent, a subbasin containing both Level I and Level II sources is initialized as SAI_I = 100% and SAI_II = 100%.

**Step 3: Propagate indices through the river network.** Computationally traverse the routing network from headwaters to the outlet. For each downstream subbasin *k*, compute SAI_I and SAI_II as drainage area-weighted averages of the indices of its directly connected upstream subbasins, as given by Eqs. (1) and (2), respectively. During the network traversal, if subbasin *k* contains a Level I alteration signal source, $\text{SAI_I}_k$ is set to 100% regardless of upstream values (e.g., SAI_I resetting at the outlet of subbasin F in Figure 1); similarly, if subbasin *k* contains a Level II signal source, $\text{SAI_II}_k$ is set to 100%. This same-level “reset” reflects the assumption that once flow passes through a signal source, the corresponding alteration becomes locally dominant at that location and downstream of it.

$$\text{SAI_I}_k = \begin{cases} 100\%, & \text{Subbasin contains Level I signal source} \\ \frac{\sum_{j=1}^{n}(\text{SAI_I}_j \times D_j)}{D_k} \times 100\%, & \text{Otherwise} \end{cases} \tag{1}$$

$$\text{SAI_II}_k = \begin{cases} 100\%, & \text{Subbasin contains Level II signal source} \\ \frac{\sum_{j=1}^{n}(\text{SAI_II}_j \times D_j)}{D_k} \times 100\%, & \text{Otherwise} \end{cases} \tag{2}$$

where *k* is the index of a given subbasin within the routing network; *n* is the total number of subbasins immediately upstream of subbasin *k* (i.e., those directly draining into subbasin *k*), indexed by *j*; *j* denotes the *j*th directly connected upstream subbasin to subbasin *k*; $D_k$ represents the total drainage area of subbasin *k*, encompassing its own area and the areas of all upstream subbasins; $\text{SAI_I}_j$ and $\text{SAI_II}_j$ represent the SAI values calculated for each directly connected upstream subbasin *j*. The resulting SAI values $\text{SAI_I}_k$ and $\text{SAI_II}_k$ are expressed as percentages.

SAI values represent the expected degree of alteration at each subbasin outlet along the routing network (Figure 1b). SAI_I ranges from 0 to 100% and reflects influence associated with confirmed (Level I) sources, whereas SAI_II ranges from 0 to 100% and reflects influence associated with potential (Level II) sources. Downstream propagation follows a drainage area-weighted mixing assumption: alteration attenuates as incremental contributing area and tributary inflows add water from natural subbasins and dilute the upstream signal. This behavior is illustrated in Figure 1b, where SAI decreases between confluences as contributing area increases, and step changes (i.e., resetting) occur at subbasins containing new sources. When a subbasin contains a new alteration source, the corresponding same-level index is reset to its source magnitude (SAI_I = 100% for Level I; SAI_II = 100% for Level II) prior to further downstream propagation. Unlike indices based on the summed storage of all upstream reservoirs, this resetting rule is designed to track the spatial imprint of alteration signals along the network, so that SAI reflects downstream propagation and dilution rather than simple cumulative accumulation of upstream infrastructure.

Furthermore, since SAI_I and SAI_II are propagated independently, it allows an upstream Level I signal to remain traceable downstream even after the network passes a Level II source, thereby avoiding artificial truncation by a weaker downstream Level II signal.

Overall, the SAI framework provides a parsimonious, network-based characterization of streamflow alteration at the subbasin scale that is readily interpretable for applications such as delineating downstream influence zones, screening hydrometric gauges for dam impacts, and supporting aquatic and ecological analyses, without requiring detailed dam or reservoir operating information. For subsequent applications, SAI_I and SAI_II values can be translated into categorical classes (e.g., natural, conditional, altered) using predefined thresholds (Section 2.5).

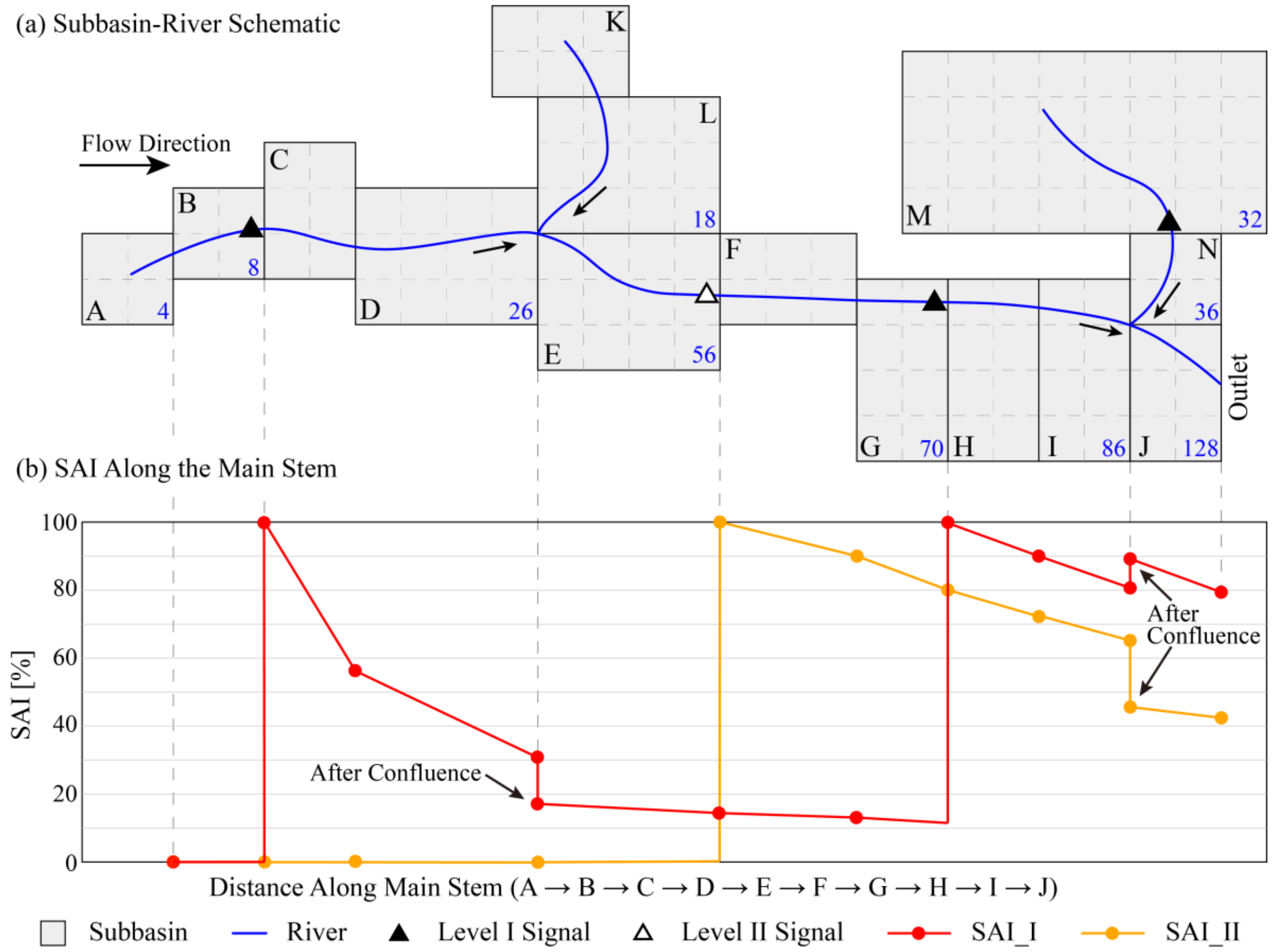

**Figure 1:** Conceptual illustration of the streamflow alteration index (SAI) propagation along a routing network. (a) A simplified subbasin-river schematic (lakes omitted for brevity) showing the main stem (A → B → C → D → E → F → G → H → I → J) with two tributaries (K → L and M → N) joining at subbasins E and J. Four alteration signal sources are located in subbasins B, E, G and M (filled triangle: Level I; open triangle: Level II). Subbasins are overlaid with equal-sized dashed grid cells to facilitate visual comparison of contributing areas, and blue numbers in the lower-right corners denote drainage area at the subbasin outlet (expressed as the number of dashed grid cells). (b) Example longitudinal profiles of SAI_I (red) and SAI_II (orange) along the main stem, illustrating (1) downstream attenuation (dilution) of alteration signals with increasing contributing area and tributary inflows, and (2) SAI resetting to source magnitude when the flow passes a new same-level signal source (e.g., SAI_I resets to 100% at the outlet of subbasin G). Note that the SAI lines denote continuous changes in SAI along the flow path, with linear attenuation assumed within each subbasin, whereas the points reflect the discrete SAI values at subbasin outlets along the flowpath. The paired points at the outlets of subbasins E and J show SAI values immediately before and after tributary confluence. The distance shown on the *x*-axis in panel (b) is approximate, since the river segments in panel (a) are schematic and not drawn to scale.

## 2.5 Validation of SAI at Hydrometric Gauges

The accuracy and reliability of SAI depend on the completeness and quality of the datasets used to identify alteration signal sources. We therefore prioritized provincial inventories over global datasets such as the GeoDAR (Wang et al., 2022) and Global Dam Watch (Lehner et al., 2024) when developing SAI v1.0 for Ontario. Although global products provide broad coverage and have undergone quality control, our province-scale data source inspection showed that global sources can miss important dams, reservoirs, and major waterpower facilities in Ontario, which would lead to spatial under-identification of alteration signals and underestimation of alteration extent.

We retained only active Water Survey of Canada (WSC) streamflow and water level gauges for validation that had at least three years of observations during 1990 to 2024. This screening removes discontinued gauges and helps ensure that the selected gauges reflect alteration conditions over recent decades, which are more consistent with the present-day influence of dams and reservoirs. Although water diversions occur in Ontario and are represented in the OLRRP network through enhanced hydrologically conditioned flow directions, diversions reflect basin-scale redistribution rather than the instream dam and reservoir alteration targeted by SAI. Gauges primarily influenced by diversions were therefore excluded, including those associated with regulated diversion works and watershed diversions recognized under Ontario water management frameworks (International Joint Commission, 2020; Ontario Ministry of Natural Resources and Forestry, 2017).

After the screening, 488 WSC gauges remained for validation. These gauges are classified in the HYDAT database as either “Regulated” or “Natural” (Water Survey of Canada, 2024). Since the RHBN gauges are a subset of the WSC gauge network, 148 RHBN gauges were included within the 488-gauge WSC set. The validation dataset therefore comprised three gauge groups: (1) 148 RHBN natural gauges, (2) 125 WSC-natural gauges outside the RHBN set, and (3) 215 WSC-regulated gauges.

RHBN gauges have been previously screened to represent virtually natural streamflow conditions with high confidence (Brimley et al., 1999; Ouarda et al., 1999; Pellerin, 2019). By contrast, according to the HYDAT metadata (see Data Availability), WSC-regulated gauges are defined as gauging stations with an upstream regulating structure, such as a dam or weir. This classification is useful, but it is generally less certain than RHBN, because WSC regulation metadata are not as complete or consistently documented. Accordingly, comparison with the WSC regulated/natural classification should be interpreted only as a broad consistency check rather than as a rigorous validation of SAI.

We validated SAI v1.0 by comparing SAI-derived conditions at hydrometric gauge locations against independent agency classifications from the WSC and the Reference Hydrometric Basin Network (RHBN). For each gauge (or subbasin), we extracted the corresponding SAI_I and SAI_II values and assigned one of three categories: a location is *natural* if SAI_I ≤ 20% and SAI_II ≤ 40%; it is *altered* if SAI_I > 20%; and it is *conditional* otherwise (i.e., SAI_I ≤ 20% but SAI_II > 40%). This three-class scheme prioritizes SAI_I for identifying altered conditions, because Level I sources represent confirmed and typically stronger regulation signals, while SAI_II is used to flag cases where alteration is plausible but more uncertain. Thus, the conditional class indicates limited confirmed alteration but elevated potential alteration influence.

These classification thresholds were selected to provide a conservative separation between natural and altered conditions. We adopted an SAI_I threshold of 20% because Level I signals represent confirmed and typically stronger alteration, which should remain limited in natural basins. This threshold is also consistent with agency instream flow assessment practice, which commonly applies pragmatic decision rules to delineate downstream impact limits for waterpower projects, including watershed-area scaling criteria such as the “5 times drainage area” rule described by Lewis et al. (2004). We adopted an SAI_II threshold of 40% because Level II signals indicate potential rather than confirmed alteration, and thus a somewhat more permissive threshold is appropriate. These thresholds are used only for screening and classification in the validation analysis and related SAI-based applications; they do not affect the underlying SAI calculations.

## 3. Data Records and Technical Validation

This section describes the Streamflow Alteration Index (SAI) v1.0 database, including the GIS layers provided and their key attributes (Section 3.1), followed by the technical validation of the database (Section 3.2).

### 3.1 SAI v1.0 Database

The SAI v1.0 database for Ontario is publicly available under the Creative Commons Attribution 4.0 International (CC BY 4.0) license at https://doi.org/10.5281/zenodo.19547932 (Shen et al., 2026). The release contains three primary data products:

- **Subbasin polygon layer (ontario_subbasin_sai_v1-0.shp):** An Esri Shapefile containing SAI values for all subbasins in the Ontario Lake and River Routing Product (OLRRP) v2 network.
- **River polyline layer (ontario_river_v2-0-1.shp):** An Esri Shapefile containing SAI values for all river reaches in the Ontario Lake and River Routing Product (OLRRP) v2 network, which is paired with the subbasin layer.
- **Alteration signal source table (ontario_alteration_signals.csv):** An ASCII (tabular) dataset describing all Level I, Level II, and Level III alteration signal sources in Ontario.

The Esri Shapefile is in the NAD 1983 Ontario MNR Lambert coordinate reference system (Lambert Conformal Conic). The datasets and their attributes are described below.

**(1) Ontario Alteration Signal Sources**

This CSV table contains all 2,475 alteration signal sources identified in Ontario and used in the SAI workflow (Table 1). Since sources were compiled from multiple agencies and inventories, original metadata and attribute structures were not uniform. We therefore retained only the core fields required for signal identification and SAI processing and harmonized key identifiers and location fields to a consistent schema.

As summarized in Table 2, fields include: the OLRRP drainage region label (Region); the source dataset indicator (SRC_obs), which records the provenance of each record (e.g., provincial dam inventories, waterpower facility datasets, and lockstations); a source identifier (Obs_NM) and, where available, a descriptive name (Gauge_nm); the associated OLRRP subbasin identifier (SubId); geographic coordinates (Lat, Lon) referenced to NAD 1983; and binary indicators (Level_I, Level_II, Level_III) that flag whether the source is treated as a confirmed, potential, or negligible alteration signal for SAI mapping.

Name-related fields are not available for all source types. Records in the signal-source table can be linked to the subbasin polygon layer via SubId (recommended) to retrieve the corresponding SAI_I/SAI_II values and additional OLRRP v2 attributes for the hosting subbasin.

**Table 2.** Attributes in the alteration signal sources table of the Streamflow Alteration Index (SAI) v1.0 database for Ontario.

| Attributes | Descriptions |
|---|---|
| Region | Drainage region of Ontario that served as subdomains in the OLRRP v2 delineation, including SW, NE, SE, NW, NC, FNE, FNC1, FNC2, and FNW. |
| SRC_obs | Source of the points. WSC: Gauge points from the Water Survey of Canada (WSC); ODI: dam points from the Ontario Dam Inventory; MNR: dam points from the Ontario MNR; WPF: waterpower facility points from the Ontario MNR; LOCKSTATION: Lockstation from the Trent-Severn Waterway, the Rideau Canal, and the Welland Canal. |
| Obs_NM | The point of interest ID as reported by agency providing the location. A unique string is created for this ID if it is not provided in the original sources to track points in the SAI database. |
| Gauge_nm | A name string for points preserved from their metadata of different sources if valid. This name |

| Attributes | Descriptions |
|---|---|
| | attribute is also a description of the point such as implying if the dam is a control dam. |
| SubId | Unique identifier of subbasins in OLRRP v2. |
| Lat/Lon | NAD 1983 geographic coordinates. |
| Level_I | Binary variable to indicate if an alteration signal source is Level I. |
| Level_II | Binary variable to indicate if an alteration signal source is Level II. |
| Level_III | Binary variable to indicate if an alteration signal source is Level III. |

**(2) Ontario Subbasin-Scale SAI**

The subbasin polygon layer delineates 245,576 subbasins across Ontario and provides SAI_I and SAI_II values (in %) for each subbasin outlet. Subbasin boundaries and network topology are employed from the Ontario Lake and River Routing Product (OLRRP) v2, with a mean subbasin area of approximately 4.3 $km^2$. Attribute fields are documented in Table 3 and include: (1) the OLRRP v2 drainage region identifier (Region); (2) the unique subbasin identifier (SubId) and its immediate downstream identifier (DowSubId), which together define the directed river network topology; and (3) subbasin-scale SAI values (SAI_I and SAI_II).

Since SubId and DowSubId follow the OLRRP v2 schema, users can link the SAI layer to the full OLRRP v2 attribute tables to retrieve additional subbasin- and reach-level descriptors (e.g., drainage area, river length, river slope, and other routing parameters) using table joins on SubId. A companion river-reach polyline layer is provided with a consistent identifier schema; however, the number of reach features is smaller than the number of subbasins, because some subbasins (e.g., headwaters) do not contain a mapped river reach.

**Table 3.** Attributes in the subbasin/river reach layer of the Streamflow Alteration Index (SAI) v1.0 database for Ontario.

| Attributes | Descriptions |
|---|---|
| Region | Drainage region of Ontario that served as subdomains in the OLRRP v2 delineation, including SW, NE, SE, NW, NC, FNE, FNC1, FNC2, and FNW. |
| SubId | Unique identifier of subbasins in OLRRP v2. |
| DowSubId | ID of downstream subbasin. |
| SAI_I | Streamflow alteration index value for Level I signal sources at the outlet of the subbasin, in percentage. |
| SAI_II | Streamflow alteration index value for Level II signal sources at the outlet of the subbasin, in percentage. |

The Streamflow Alteration Index (SAI) v1.0 database provides a province-wide, subbasin-scale representation of instream alteration by propagating point-based alteration signal sources downstream through the connected OLRRP v2 network. The resulting spatial patterns of SAI_I (0–100%) and SAI_II (0–100%) are mapped across Ontario in Figure 2, revealing pronounced heterogeneity in streamflow alteration within and among the nine drainage regions. To illustrate these propagation patterns more clearly at the regional scale, Figure 3 presents higher-resolution maps for the Southeast (SE) region, where the spatial imprint of individual signal sources and their downstream influence can be seen more clearly along connected river networks.

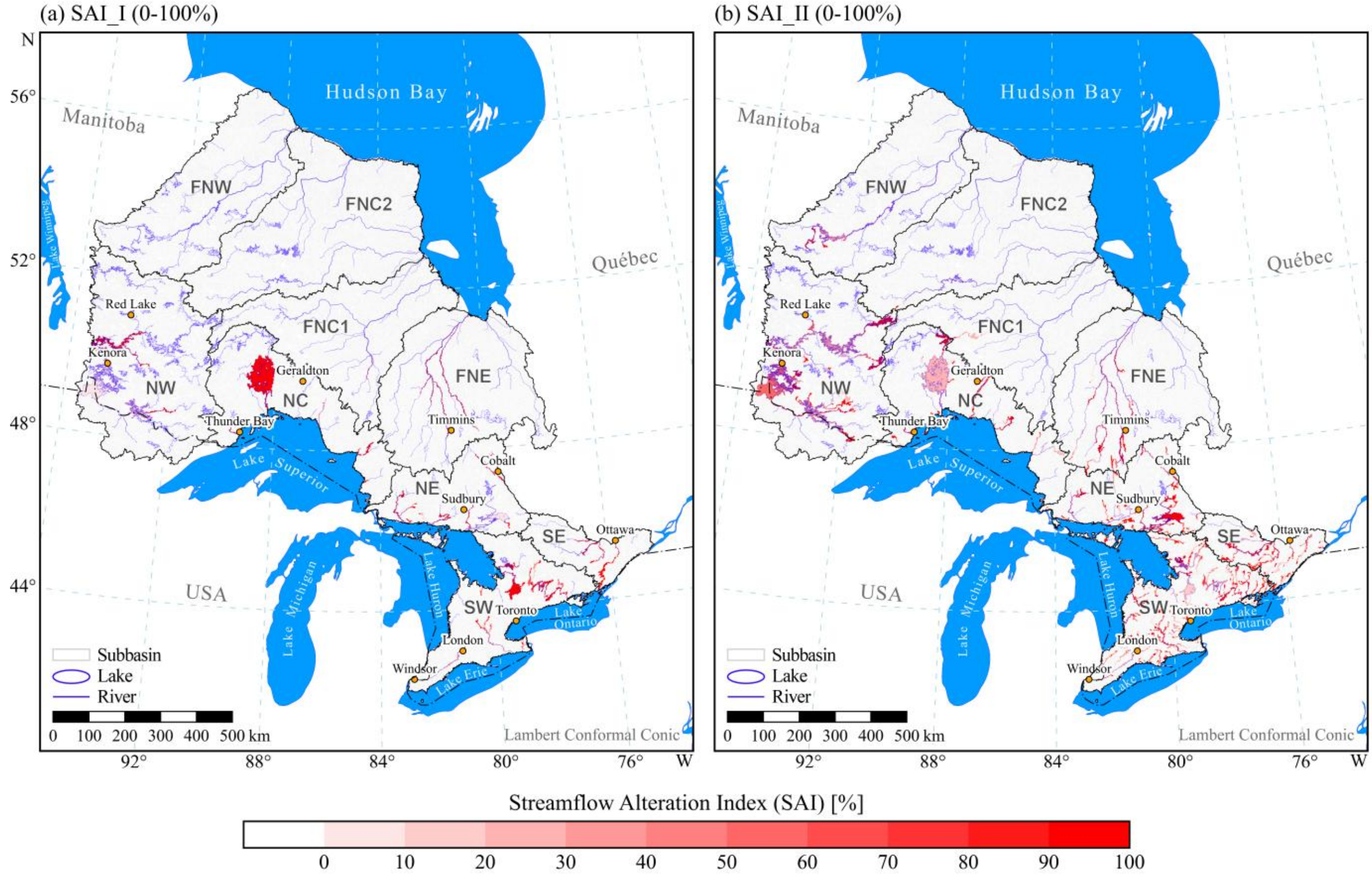

**Figure 2:** Spatial distribution of subbasin-scale (a) SAI_I and (b) SAI_II across Ontario. Subbasin boundaries (gray outlines; fill colors indicate SAI magnitude), rivers (blue lines; thicker lines indicate higher Strahler stream order), and lakes (blue outlined polygons; only lakes with area > 100 km$^2$ are shown) are from the Ontario Lake and River Routing Product Version 2 (OLRRP v2). Ontario is partitioned into nine independent drainage regions: Southwest (SW), Southeast (SE), Northeast (NE), North Central (NC), Northwest (NW), Far Northeast (FNE), Far North Central 1 (FNC1), Far North Central 2 (FNC2), and Far Northwest (FNW).

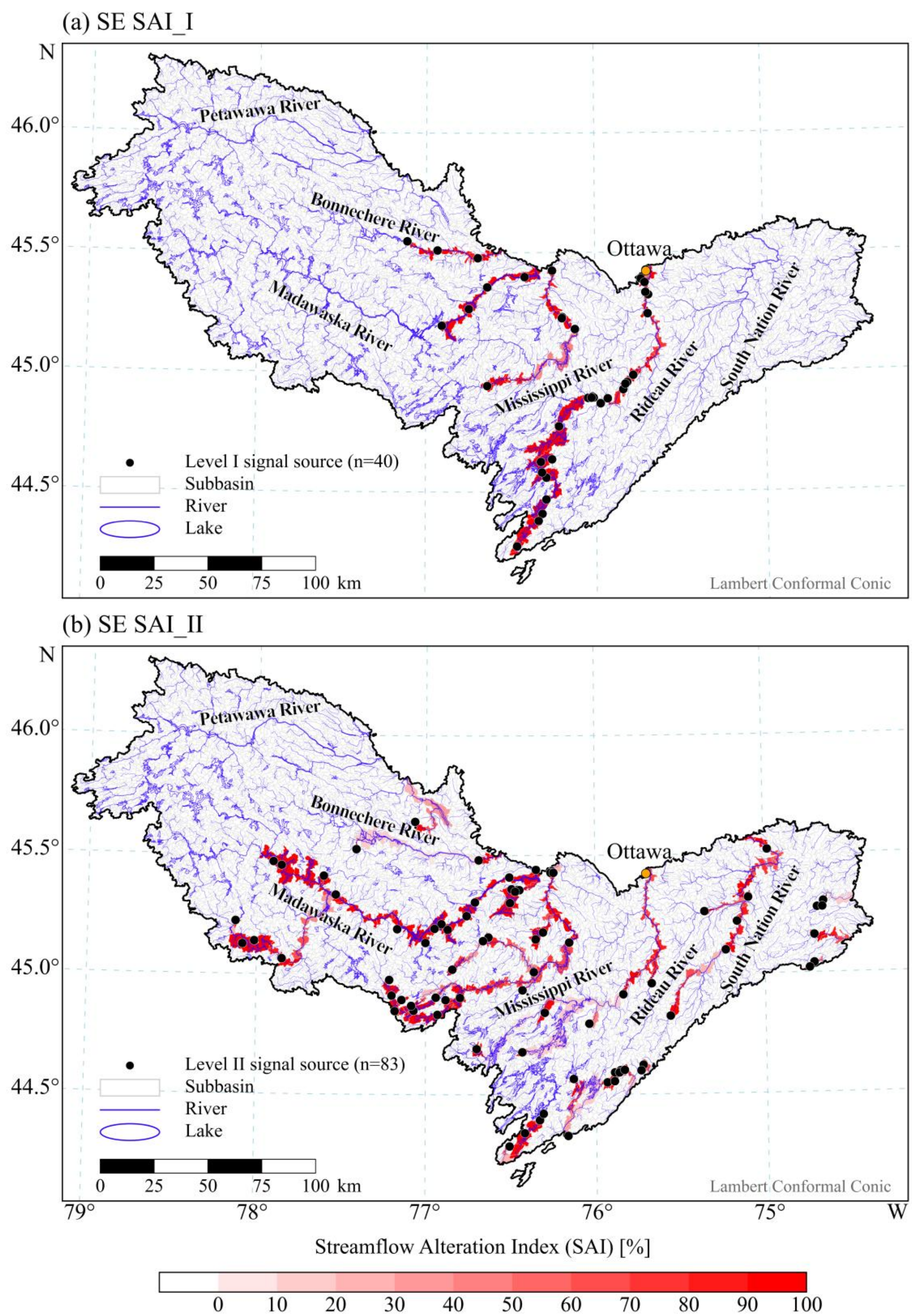

**Figure 3:** Higher-resolution maps of subbasin-scale (a) SAI_I and (b) SAI_II for the Southeast (SE) region of Ontario. Black points indicate Level I and Level II signal sources for SAI_I and SAI_II, respectively. Subbasin boundaries (gray outlines; fill colors indicate SAI magnitude), rivers (blue lines), and lakes (blue outlined polygons; all lakes are shown) are from the Ontario Lake and River Routing Product Version 2 (OLRRP v2).

Interactive high-resolution regional SAI maps are also available through the archived dataset, enabling users to explore local propagation patterns beyond the province-wide overview shown in Figure 2. The SE example in Figure 3 illustrates the finer spatial detail available in these regional products.

### 3.2 Validation of SAI v1.0 Database

This evaluation is intended to assess the usefulness of SAI as a screening dataset for instream alteration, rather than to verify the exact magnitude of hydrologic alteration at individual sites. To validate the database, we first extracted SAI_I and SAI_II values for the OLRRP v2 subbasins containing hydrometric gauges, then used these to assign hydrometric gauges a SAI-based gauge class (natural, conditional, altered) using the thresholds defined in Section 2.5. We compared the resulting SAI classifications against two independent reference datasets: Reference Hydrometric Basin Network (RHBN) gauges and Water Survey of Canada (WSC) gauge classifications (Section 2.5). Validation gauges were organized into three reference groups: (1) 148 RHBN natural gauges, (2) 125 WSC-natural gauges outside the RHBN subset, and (3) 215 WSC-regulated gauges. Results are summarized in Figure 4.

Across gauges identified as natural by the reference datasets, SAI showed strong agreement. For RHBN natural gauges (n = 148), 86% were classified as natural by SAI and 14% as conditional, with no gauges classified as altered (Figure 4a). The 21 RHBN gauges assigned to the conditional class indicate that there may be some inconsistency with this high-confidence natural benchmark, which highlights that the present validation is necessarily approximate rather than exact. For WSC-natural gauges outside the RHBN subset (n = 125), 86% were classified as natural, 12% as conditional, and 2% as altered (Figure 4b). Agreement was lower, as expected, for WSC-regulated gauges (n = 215), for which SAI classified 26% as altered, 40% as conditional, and 34% as natural (Figure 4c). At minimum, this result suggests that the WSC regulation flag is relatively conservative and may include basins that are only slightly altered or uncertain with respect to the type of instream alteration represented by SAI. Overall, the pattern of disagreement in Figure 4 is consistent with expectations: the least disagreement occurs for the highest-confidence reference dataset (RHBN), whereas the greatest disagreement occurs for the lower-confidence WSC-regulated class.

Disagreements between SAI-based classes and agency labels are expected for several reasons: (1) The reference classifications are not designed to isolate instream development impacts alone: RHBN screening may reflect broader watershed pressures beyond instream infrastructure, and the basis for WSC "natural" and "regulated" labels is not consistently reported at a level that enables one-to-one mapping to SAI; (2) SAI explicitly represents uncertainty through SAI_I and SAI_II by incorporating both confirmed and potential alteration signal sources, which can yield conditional classifications where evidence is incomplete; and (3) The alteration signal source inventory may be incomplete for some locations, which can lead to underestimation of alteration where regulation exists but is not represented in available datasets. More broadly, the mismatches in Figure 4 should be interpreted in light of uncertainty in both the reference labels and the SAI target classes, rather than as definitive classification errors in either dataset. These considerations motivate interpreting SAI as a practical screening tool that provides consistent, spatially comprehensive classifications, while allowing users to incorporate additional local knowledge (e.g., hydrograph inspection or consultation of detailed upstream dam and reservoir information) when agency labels and SAI classes disagree.

Despite these limitations, Figure 4 indicates that SAI v1.0 provides a practical provincial-scale screening tool. It preserves a large majority of gauges labelled natural by RHBN and WSC, while partitioning WSC-regulated gauges into altered, conditional, and natural subsets that can be prioritized for follow-up assessment.

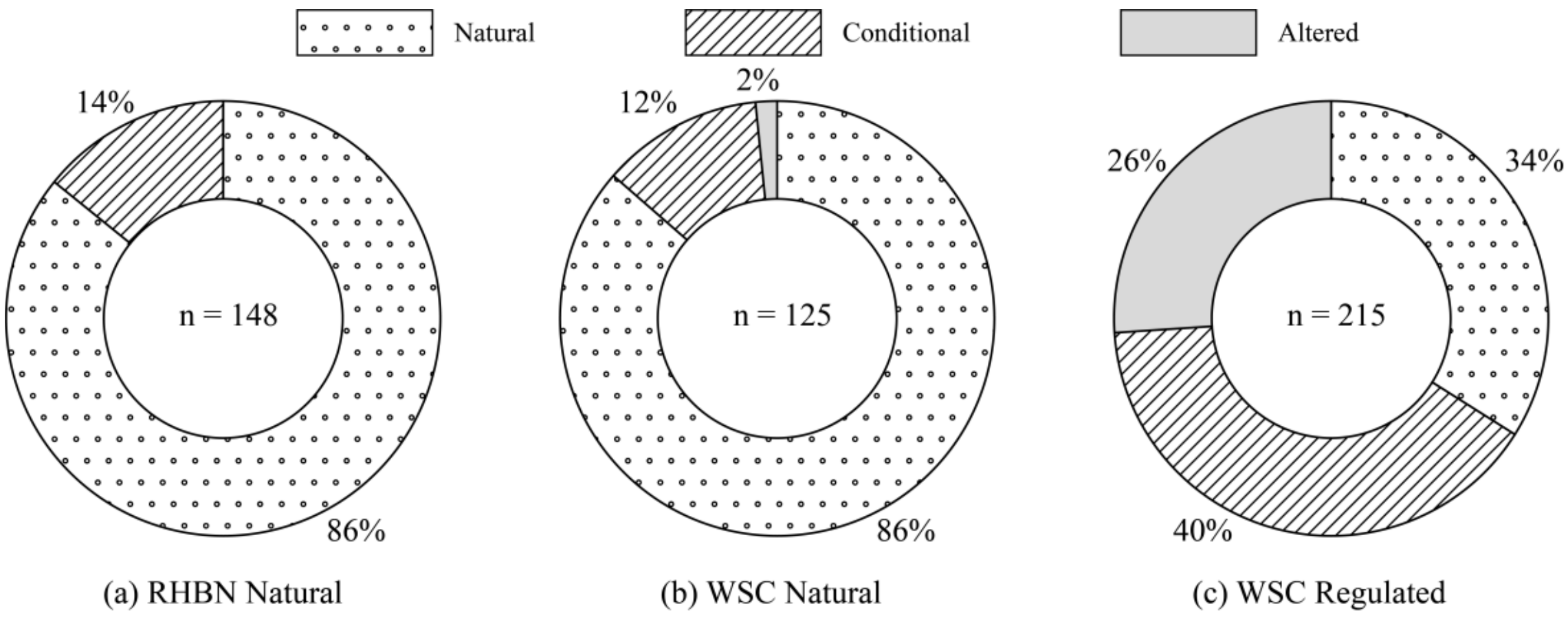

**Figure 4.** Summary of the Streamflow Alteration Index (SAI) validation results for three reference gauge groups: (a) RHBN natural gauges, (b) WSC-natural (non-RHBN) gauges, and (c) WSC-regulated gauges. Donut charts show the proportion of gauges assigned to each SAI class (natural, conditional, altered) based on the classification rules in Section 2.5. Natural is defined as SAI_I ⩽ 20% and SAI_II ⩽ 40%; altered is defined as SAI_I > 20%; all remaining cases are classified as conditional. The number of gauges in each reference group (n) is shown at the center of each chart, and percentages indicate the share of gauges in each SAI-based class.

## 4. Discussion and Limitations

The Streamflow Alteration Index (SAI) is not intended to quantify the precise magnitude of streamflow alteration at a specific location—for example, the volume of water stored in reservoirs, operating releases, or the detailed changes in hydrograph magnitude and timing relative to natural conditions. Instead, SAI is designed to provide a spatially continuous, network-based screening metric with an approximate degree of alteration computed based on the percentage of upstream drainage area routed through an upstream alteration source. Since SAI does not require site-specific observations or dam operating rules, it is readily applicable at large spatial scales. Therefore, SAI values should be interpreted as indicating the presence and potential influence of instream alteration rather than its absolute hydrologic magnitude. For example, a large main stem subbasin downstream of a major dam may exhibit an SAI value comparable to that of a smaller tributary affected by a smaller structure, even though the underlying volumes of altered flow may substantially differ.

In practice, SAI helps delineate the downstream footprint of instream alteration and identify hydrometric gauges that may be influenced by upstream dams and reservoirs. This capability reduces uncertainty when screening hydrometric gauges, particularly for large monitoring networks where "natural" versus "regulated" labels are not always accompanied by transparent, consistent criteria. As demonstrated in Section 3.2, SAI aligns strongly with RHBN natural gauges and provides a consistent, spatially comprehensive complement to agency classifications when selecting calibration and validation data for large-domain hydrologic modeling.

A key limitation is that SAI v1.0 targets instream alteration signals and does not explicitly represent landscape-driven alterations (e.g., land-use change, drainage modifications, agriculture, or urbanization). As a result, gauges classified as natural by SAI may still be affected by non-structural watershed pressures that are outside the scope of the current signal inventory. Hydrologic modelers should therefore treat SAI as an initial filter and incorporate additional information, such as land-use datasets, water withdrawals, or hydrograph-based diagnostics, when evaluating gauge suitability for specific applications. Even so, SAI can significantly reduce the initial workload by removing gauges with clear evidence of upstream instream alteration.

Another limitation is that downstream attenuation in SAI is represented using a drainage area-weighted mixing assumption only. In reality, subbasins with similar drainage area may differ in their effective attenuation capacity because of lakes, wetlands, channel storage, and routing characteristics. In particular, subbasins containing lakes may exert stronger flow-buffering and hydrograph-smoothing effects than lake-free subbasins of comparable size (Hudson et al., 2021). As a result, the current formulation should be viewed as a first-order approximation of downstream dilution rather than a full representation of hydrologic attenuation processes.

A further methodological simplification is that Level I and Level II signals are propagated independently. As a result, a downstream Level II source does not attenuate or replace an upstream Level I signal. This behavior was intentional, because Level I signals dominate the identification of altered sites in the current screening framework, and preserving their traceability helps avoid artificial suppression by more uncertain downstream Level II sources. Therefore, this simplification is expected to have limited effect on the final altered–natural–conditional classification, although it may not fully represent more complex interactions among sequential source types along the same river path.

Moreover, incomplete or uncertain inventories of dams and waterpower facilities can lead to underestimation of alteration where infrastructure is missing from available datasets. We therefore recommend using SAI v1.0 as a preliminary reference for gauge screening and for interpreting model performance in Ontario, followed by targeted basin-specific review where needed, especially where SAI_II is elevated, which may indicate potential alteration influence with higher uncertainty.

The successful validation and demonstrated utility of SAI v1.0 in Ontario suggest strong potential for extending the methodology to broader spatial domains. Ongoing advances in large-sample dam and reservoir inventories, together with the increasing availability of high-resolution hydrographic routing frameworks at national, continental, and global scales, make it feasible to implement the SAI framework beyond Ontario and to develop network-based screening products that support more consistent assessment of instream alteration for hydrologic and ecological applications at large scales.

## 5. Conclusions

We developed the streamflow alteration index (SAI) v1.0 database for Ontario to provide a spatially continuous, subbasin-scale screening product that maps the downstream influence of instream alteration signal sources across a connected hydrographic network. The database includes paired indices (SAI_I and SAI_II) that represent two different levels of alteration influence based on confirmed and potential sources, enabling consistent classification of streamflow at subbasin outlets (and thus hydrometric gauge locations) as natural, conditional, or altered. Our SAI database and suite of online maps across Ontario enable water resources researchers and practitioners to quickly assess the potential degree of streamflow alteration for any point on the surface water network in Ontario, while accounting for differences in source confidence.

Validation against independent reference classifications indicates that SAI classified 86% of RHBN natural gauges and 86% of WSC-natural gauges as natural, while 74% of WSC-regulated gauges were classified as conditional or natural.

Overall, SAI v1.0 offers a practical, network-based tool for large-domain hydrologic and ecological applications, including gauge screening, model calibration/validation support, and basin-scale prioritization of locations potentially affected by upstream alteration. Future work will focus on improving alteration source inventories, extending the framework to larger scales (e.g., national applications across Canada), and integrating complementary information to better represent non-structural watershed alterations.

## Data Availability

The Streamflow Alteration Index (SAI) v1.0 database developed in this study, including the subbasin polygon layer, river polyline layer, alteration signal source table, and accompanying documentation, is archived on Zenodo under the Creative Commons Attribution 4.0 International (CC BY 4.0) licence: https://doi.org/10.5281/zenodo.19547932 (Shen et al., 2026). The dataset should be cited as follows:

The Zenodo metadata record is publicly accessible. During manuscript review, the data files are available to editors and reviewers through the anonymous access link below. The files will be made openly available upon publication.

https://zenodo.org/records/19547932?token=eyJhbGciOiJIUzUxMiJ9.eyJpZCI6IjhiZTE0YmE5LTUxOTQtNGNiNS04MzA4LTVkNTBhZGZlNjNhNCIsImRhdGEiOnt9LCJyYW5kb20iOiI2ODUyNTIyNjZlYTIxMWIyNzM3MDc5MGMzN2VkOTgxZiJ9.mcylZOwiZ3MMBAniEtaGmkYl9MuIiK3vf70rhgKtsLZBNNc-PgV9yWyUhLC5p4tubkLiWTT5kn7ChlbgXRKgwQ

An interactive online SAI map for Ontario will be released upon publication. For manuscript review, an example interactive map for the Southeast (SE) drainage region is available below.

https://www.arcgis.com/apps/mapviewer/index.html?layers=b58a3e73c15d4f0ca61e1f5c62a2acde

All input datasets used in this study are publicly available from their original sources:

- Ontario Lake and River Routing Product, Version 2 (OLRRP v2): available through Shen et al. (2024). OLRRP website at: https://uwaterloo-olrrp.shinyapps.io/OLRRP-V2/
- Ontario Dam Inventory (ODI): available from the Ontario Ministry of Natural Resources (MNR). Download at: https://geohub.lio.gov.on.ca/maps/mnrf::ontario-dam-inventory/about
- Hydrometric gauge data (Water Survey of Canada, WSC): available from the HYDAT database. Download at: https://collaboration.cmc.ec.gc.ca/cmc/hydrometrics/www/ Metadata file: https://collaboration.cmc.ec.gc.ca/cmc/hydrometrics/www/ECDataExplorer_EN.pdf
- Reference Hydrometric Basin Network (RHBN): documentation and data access available at: https://www.canada.ca/en/environment-climate-change/services/water-overview/quantity/monitoring/survey/data-products-services/reference-hydrometric-basin-network.html
- Trent-Severn Waterway National Historic Site: https://parks.canada.ca/lhn-nhs/on/trentsevern/visit/posteeclusage-lockstation
- Rideau Canal National Historic Site: https://parks.canada.ca/lhn-nhs/on/rideau/visit/posteeclusage-lockstation
- Welland Canal locks: https://greatlakes-seaway.com/en/the-seaway/our-locks-and-channels/

## Code Availability

The scripts used to calculate and propagate SAI values through the OLRRP v2 routing network are archived in a public GitHub repository below. These scripts reproduce the core SAI calculation workflow used to generate the SAI v1.0 database.
https://github.com/hongren-shen/streamflow-alteration-index/tree/main

## Author Contributions

**Hongren Shen** led the study, including data compilation and processing, methodological development, SAI calculation, analysis, figure preparation and visualization, and manuscript writing. **Bryan A. Tolson** and **James R. Craig** contributed to methodological design, interpretation of results, figure visualization, and manuscript revision. **Robert A. Metcalfe** and **Jamie Luce** contributed data resources, interpretation of results, and manuscript revision. **Jonathan Romero-Cuellar** contributed to interpretation of results and manuscript revision. All authors reviewed and approved the final manuscript.

## Competing interests

The authors declare that they have no conflict of interest.

**Disclaimer**

Copernicus Publications remains neutral with regard to jurisdictional claims made in the text, published maps, institutional affiliations, or any other geographical representation in this paper. While Copernicus Publications makes every effort to include appropriate place names, the final responsibility lies with the authors. Views expressed in the text are those of the authors and do not necessarily reflect the views of the publisher.

**Acknowledgements**

The authors thank the Ontario Ministry of Natural Resources, the Water Survey of Canada, Parks Canada, and other organizations that maintain and provide access to the datasets used in this study.

**Financial Support**

This research was supported by a 2022–2026 Collaborative Research Agreement between the Ontario Ministry of Natural Resources and the University of Waterloo (Principal Investigator Dr. Bryan Tolson) and this funding supported Dr. Shen.